\documentclass[copyright]{eptcs}
\providecommand{\event}{UITP 2014} 
\usepackage{breakurl}             
\usepackage{amssymb,amsmath}
\usepackage{saoithin}
\usepackage{graphicx}

\title{\UTP2: Higher-Order Equational Reasoning by Pointing}
\author{
  Andrew Butterfield\thanks{This work was supported, in part, by Science Foundation Ireland grant 10/CE/I1855}
  \institute{
     School of Computer Science and Statitics\\
     Trinity College Dublin\\
     Ireland}
  \email{Andrew.Butterfield@scss.tcd.ie}
}
\def\titlerunning{UTP2: Reasoning by Pointing}
\def\authorrunning{A. Butterfield}
\begin{document}
\maketitle

\begin{abstract}
We describe a prototype theorem prover, \UTP2, developed to match the style
of hand-written proof work in the Unifying Theories of Programming semantical
framework. This is based on alphabetised predicates in a 2nd-order logic,
with a strong emphasis on equational reasoning.
We present here an overview of the user-interface of this prover,
which was developed from the outset using a point-and-click approach.
We contrast this with the command-line paradigm that continues to dominate
the mainstream theorem provers,
and raises the question: can we have the best of both worlds?
\end{abstract}

\section{Introduction}\label{sec:intro}

Unifying Theories of Programming (UTP) \cite{UTP-book},
is a framework that uses alphabetised predicates to define language
semantics in a relational calculus style, in a way that facilitates
the unification of otherwise disjoint semantic theories,
either by merging them, or using special linking predicates
that form a Galois connection. The framework is designed
to cover the spectrum from abstract specifications
all the way down to near-machine level descriptions,
and as a consequence the notion of refinement plays a key role.

We are doing foundational work in the UTP \cite{UTP-book},
which requires formal reasoning with not only predicates,
but also predicate transformers: $\RR3(P)$ $\defs$ $\Skip \cond{wait} P$
and predicates over predicates: $P = \RR3(P)$.
We also need to use recursion at the predicate level:
$ P \defs \mu Q \bullet F(Q)$,
as well as partially-defined expressions:
$s \le s \cat (tr'-tr) \equiv  tr \le tr'$.
The logic being used is therefore semi-classical
(two-valued logic, but expressions may be undefined)
and of least 2nd-order.
In addition, tool support for foundational work in UTP requires the ability
to easily describe new language constructs,
which can themselves be treated just like predicates,
in keeping with the ``programs are predicates''
philosophy \cite{predprog} of UTP.
In \cite{conf/utp/Butterfield10}
we gave an overview of the Unifying Theories of Programming Theorem Prover
(\UTP2)
that we are developing to support such theory development work%
\footnote{%
In that paper it was called \STHN, but the name has since changed to \UTP2
}%
.
The prover is an interactive tool, with a graphical user-interface,
designed to make it easy to define a UTP theory and to experiment
and perform the key foundational proofs.
The motivation for developing this tool,
rather than using an existing one,
has been discussed in some detail
in \cite{conf/utp/Butterfield10}, but key elements will be reprised here.
The logical and  technical underpinning was further elaborated
upon in \cite{conf/utp/Butterfield12},
which described as being an adapted and generalised version of
the equational reasoning system developed by Tourlakis \cite{journals/logcom/Tourlakis01},
itself inspired by the equational logic of David Gries and his colleagues \cite{gries.93}.

In this paper, we describe how the user interacts
with this theorem prover,
that was developed,
\emph{from the outset},
with the proof and reasoning styles typically used in UTP research and published work.

The key emphasis in development was to use window-based GUI techniques early on as the primary
mode of interaction, in stark contrast to most modern interactive theorem provers
that have essentially a command-line interface (most HOL flavours, CoQ, PVS, \ldots) sometimes wrapped with a
elaborate interface built on top of a highly configurable text editor
(e.g., Proof General on Emacs, new Isabelle/HOL interface on top of jEdit).

\section{Motivation}\label{sec:motivation}

There are a lot of theorem provers in existence,
of which the most prominent feature in \cite{conf/tphol/2006provers}.
Of these, the most obvious candidates for consideration for UTP prover support
are Isabelle/HOL\cite{books/sp/NipkowPW02},
PVS\cite{conf/fmcad/Shankar96},
and CoQ \cite{bk:Coq'Art:04}.
They are powerful, well-supported,
with decades of development experience
and large active user communities.
They all support higher-order logic of some form, with a command-line interface,
typically based around tactics of some form. All three require functions to be total,
but support some kind of mechanism for handling partial functions
(e.g. dependent types in PVS).
Their reasoning frameworks are based on some form of sequent calculus,
and do not support equational reasoning in a native fashion.

There has been work done on improving the user interfaces
of theorem provers of this kind.
An interesting example was ``proof by pointing'' \cite{conf/tacs/BertotKT94}
for CoQ which allowed the user to select a subterm,
whereupon it would generate and apply a tactic based on the subterm's top-level
operator.
Whilst proof-by-pointing is not supported in more recent versions of CoQ,
it has been incorporated into ``Proof General'' \cite{conf/tacas/Aspinall00},
a general purpose user interface for theorem provers, built on top of Emacs.
It supports Isabelle and Coq, among others,
and is basically a proof-script management system.
In essence it supports the command-line tactics of the provers,
allowing the user to edit proof scripts at will,
whilst maintaining prover consistency behind the scenes.
Other explorations in this area include
INKA \cite{Hutter96}, Lovely OMEGA \cite{oai:CiteSeerXPSU:10.1.1.42.1864},
Window inference \cite{Staples95},
Generalized Rewriting in Type Theory \cite{journals/eik/Basin94},
The CoRe Calculus, \cite{conf/cade/Autexier05}
and the Jape Theorem Proving framework (\url{http://japeforall.org.uk/}.)
Of the above, \cite{conf/cade/Autexier05},\cite{journals/eik/Basin94} seems designed
to support equational reasoning, but lack any notion of a GUI. In \cite{Hutter96} and \cite{oai:CiteSeerXPSU:10.1.1.42.1864} we have GUIs, but the logic/proof style is tree based.
The window inference work \cite{Staples95} has a notion of ``focus'' similar to ours,
but has no GUI, and while capable of handling equational rewrites seems to be more general.
Jape has a GUI and facilities to encode logics, but again is deduction-biased, and has no easy
way to extend the language.

\section{Interaction}\label{sec:interact}

We shall illustrate \UTP2's use by walking through a simple proof,
from a theory of sets, regarding the commutativity of set intersection.

\noindent
We start by launching the theorem prover, and we assume that some theories have been
preloaded: \texttt{Sets}%
\footnote{The \texttt{\$0} suffix is a version number}%
, \texttt{Equality}, \texttt{Logic} and \texttt{\_ROOT} (a base theory always present).
All theories have access to definitions and laws from lower theories.

\includegraphics[scale=0.5]{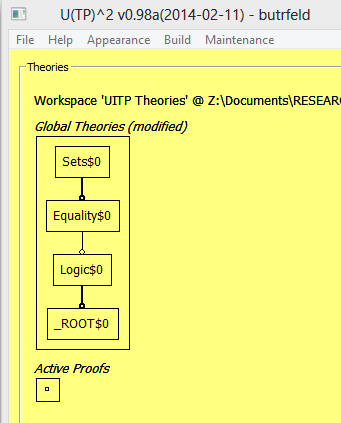}

\noindent
If we double-click on the \texttt{Sets} box,
a window opens up showing the ``Laws'' of the Set theory.
Laws have names, a ``provenance'' indicator, side-conditioning,
and their defining schema (a predicate).

\includegraphics[scale=0.5]{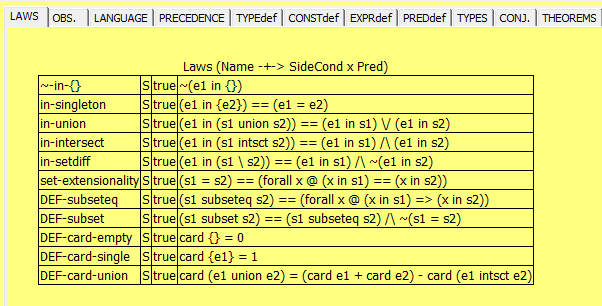}

\noindent
Clicking on the ``CONJ.'' tab shows some conveniently preloaded conjectures,
which have yet to be proven.

\includegraphics[scale=0.5]{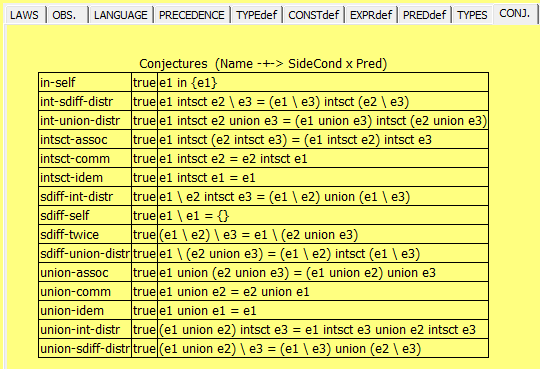}

\noindent
Double-clicking on the \texttt{intsct-comm} row (5th) opens up a proof window,
and we use its setup menu to select the ``Reduce'' strategy,
which attempts to transform the goal predicate into TRUE.
Other strategies, depending on the goal structure include: ``left-to-right'',
for equality/equivalence conjectures, that
converts the lefthand side until equal to the righthand side,
or ``reduce-both'' which tries to transform both sides into some common form.

In the proof window we have the goal and side-conditions displayed,
and there is some material about heuristics we ignore in this paper.
The lower half of the proof window displays the ``TARGET'',
determined by the goal and the chosen strategy.
Some context information is also shown, the most important being the free variables,
and the type, in this case, of each side of the equality.
We see the starting goal at the bottom, in bold and underlined%
\footnote{The ``Matches'' subwindow will not be discussed here}%
:

\includegraphics[scale=0.5]{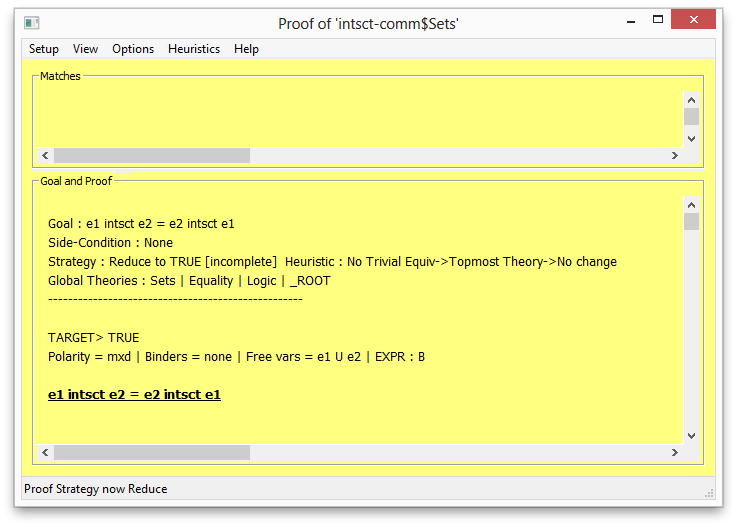}

\noindent
If we right-click anywhere in the ``Goal and Proof'' subwindow,
then a menu of laws applicable to the goal pops-up.
In effect the goal was matched against all the laws present in the
\texttt{Sets}, \texttt{Equality}, \texttt{Logic} and \texttt{\_ROOT} theories, the successful matches were then ranked
(by various user-selectable heuristics), the top twenty chosen, then applied to the
goal to show the result, and presented in the menu.

\includegraphics[scale=0.5]{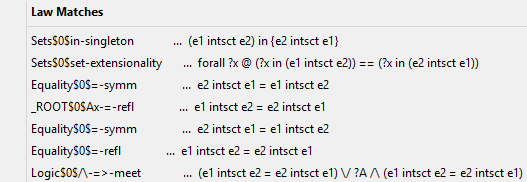}

\noindent
If we pick the second, ``set-extensionality'', as it has new variables not present in the goal, (e.g. \texttt{?x}), we are asked to supply instances for these, with a reasonable default being offered.
This feature is not obviously useful in this example (except if \texttt{x} was present elsewhere) but comes in handy when matching the rhs of a law like $A \lor (A \land B) \equiv A$, to get the rhs,
in which we are free to instantiate $B$ as we see fit.

  \includegraphics[scale=0.5]{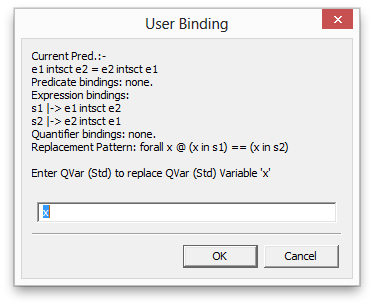}

\noindent
If we go with the default suggestion, then we obtain the following proof state:

  \includegraphics[scale=0.5]{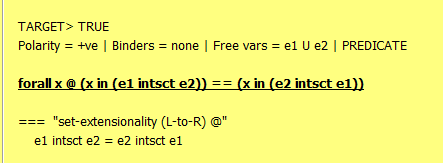}

\noindent
We can use arrow-keys to move around the goal, changing the proof ``focus''.
If we go ``down'' twice, we focus in on the first set membership assertion.
It is worth noting that the line above records that the focus is on an expression (EXPR)
of type boolean (\texttt{B}). \UTP2\ has a on-the-fly type inference algorithm that runs
every time the focus changes%
\footnote{speed has never been a problem with this}%
, and is used by the law matching algorithm to avoid spurious matches.
We avoid lots of explicit type annotations, preferring to deal with such issues
behind the scenes. This is of course in keeping with the general traditional UTP approach
to theorem development.

\includegraphics[scale=0.5]{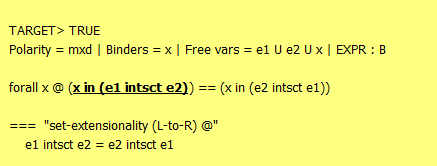}

\noindent
Right-clicking now leads to laws relevant to the focus:

\includegraphics[scale=0.5]{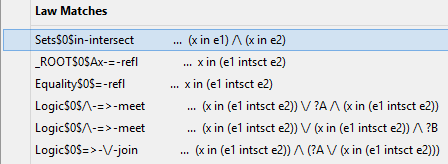}

\noindent
If we pick the first option, then we get a conjunction of simpler membership statements.

\includegraphics[scale=0.5]{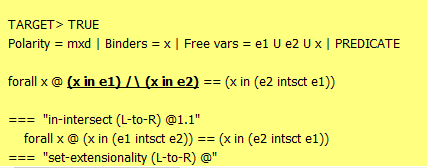}

\noindent
Moving to the righthand side of the equality, we can apply the same \texttt{in-intersect} law,
then apply the commutativity of conjunction, pull back out and we get
instances of the reflexivity of equals. Finally we get rid of a vacuous quantifier,
so resulting in the goal \texttt{TRUE}, and \UTP2\ proclaims!

\includegraphics[scale=0.5]{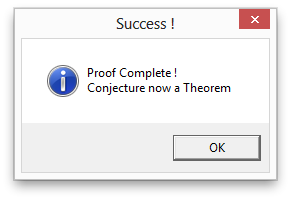}

\noindent
Examining the ``THEOREMS'' tab in the Set theory window shows our new theorem.

\includegraphics[scale=0.5]{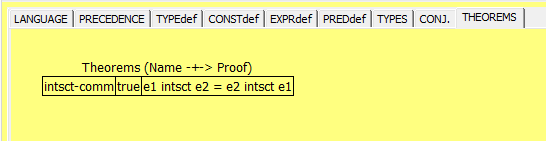}

\noindent
Right-clicking on it gives another pop-up menu of interesting things to do with it.

  \includegraphics[scale=0.5]{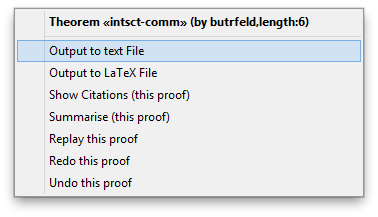}

\noindent
We render a simple text version of the resulting proof:
\begin{verbatim}
Complete Proof for 'Sets$intsct-comm
Goal : e1 intsct e2 = e2 intsct e1
Strategy: Reduce to TRUE

     e1 intsct e2 = e2 intsct e1
 ===   " set-extensionality (L-to-R) @ "
     forall x @ (x in (e1 intsct e2)) == (x in (e2 intsct e1))
 ===   " in-intersect (L-to-R) @1.1 "
     forall x @ (x in e1) /\ (x in e2) == (x in (e2 intsct e1))
 ===   " in-intersect (L-to-R) @1.2 "
     forall x @ (x in e1) /\ (x in e2) == (x in e2) /\ (x in e1)
 ===   " /\-comm (R-to-L) @1.2 "
     forall x @ (x in e1) /\ (x in e2) == (x in e1) /\ (x in e2)
 ===   " Ax-==-id (R-to-L) @1 "
     forall x @ TRUE
 ===   " forall-vac (L-to-R) @ "
     TRUE
\end{verbatim}

\noindent
We have only skimmed over the interactive proof features of \UTP2 here.
Others include
\begin{itemize}
  \item keyboard shortcuts to apply built-in procedures to the focus,
    e.g., convert to disjunctive normal form
  \item
    a clickable help feature in the proof window
  \item strategies to support inductive proofs
  \item
    all tables in each tab of each theory can be edited, with entries added
    or deleted --- even laws!
  \item
    Some of the tabs, (``OBS.'',``LANGUAGE'',``PRECEDENCE'') have
    tables that support user definitions of languages.
    See \cite{conf/utp/Butterfield12} for further details.
\end{itemize}

\section{Discussion}

Proofs done with \UTP2 are, in our opinion, more ``open'',
in that we can easily see the steps and laws used in a proof, in an equational style.
A consequence of this is readily seen when we consider the students taking the
Formal Methods course offered at Trinity College Dublin, that focusses on the
UTP, and uses \UTP2 for part of the classwork.
The feedback obtained from these students shows clearly that
(i) the learning curve to get good at \UTP2 proofs is fairly shallow---they almost never
get ``stuck'', once a few tricks are shown---experimentation is easy;
(ii) their concerns are regarding improvement to the GUI itself,
either in terms of how it looks, or having the flexibility to define their own keyboard
shortcuts.
A key feature that reduces the learning curve is the ability
of the prover to suggest possible next steps, by doing advance pattern-matching
and instantiation.

Proofs in CoQ or Isabelle/HOL are, again in this authors words, more ``procedural'',
and ``opaque'', but definitely more powerful. The disadvantage is that the learning
curve is much steeper, particularly when early success it obtained by tactics like
\texttt{auto}, \texttt{simp} or \texttt{sledgehammer}.
When these fail, the best approach is not so clear to the beginner.
However, there is undeniable power once that learning curve has been climbed.

\UTP2 was really developed to assist in the development of new semantic theories
within the UTP framework. Others have also put effort into doing this for UTP
using both ProofPowerZ\cite{conf/utp/ZC08}
and Isabelle/HOL\cite{conf/vstte/FeliachiGW12,conf/utp/FZW14}.
The price they pay is having to recast material in the ProofPower/HL style.
The benefit they tap into is the power of their proof engines.

The key questions raised here are:
\begin{itemize}
  \item Should point-n-click GUIs be added to existing provers?
  \item To what extent are front-ends like Proof-General or jEdit
  are step in this direction?
  \item
    Should more attention be paid to developing equational reasoning approaches?
  \item
  Can the \UTP2 front-end be fruitfully turned into a wrapper around Isabelle/HOL say?
  \item
  Should it use Isabelle/HOL as a way to check its proofs
  (would save trying to develop a small safe LCF-style kernel for \UTP2) ?
  \item
   Can we envisage proofs been done using gestures on a tablet?
\end{itemize}
Very recent work, presented as Tutorial 2 at FM2014 in Singapore, by Jim Woodcock, Simon Foster
and Frank Zeyda of the University of York, showed an encoding of UTP and some key theories
into Isabelle/HOL. One the negative side, they had to employ further nested quotation schemes,
but on the positive side, they used Isar in such a way that it may be relatively easy to use
Isabelle/HOL to check proof steps made by \UTP2. We hope to explore this connection in the near future.

\subsection{Obtaining Code}

\UTP2 is written in Haskell using the wxHaskell GUI library,
and is available open-source, currently under a GPL v2 license,
from \url{https://bitbucket.org/andrewbutterfield/saoithin}.
The screenshots in this paper were produced using version 0.98a.

\bibliographystyle{eptcs}
\bibliography{UITP2014-EPTCS-MAIN}

\appendix


%
%
%

\end{document}